\documentclass[twocolumn,showpacs,showkeywords,prb]{revtex4}
\usepackage{graphicx}
\usepackage{dcolumn}
\usepackage{amsmath}
\usepackage{latexsym}

\begin{document}
\title
{Noncommutative quantum mechanics of a test particle under linearized\\ gravitational waves}
\author{Anirban Saha}
\email{anirban@iucaa.ernet.in}
\affiliation{Department of Physics and Astrophysics, 
West Bengal State University,\\
Barasat, North 24 Paraganas,\\ West Bengal, India\\ }
\affiliation{Visiting Associate in Inter University Centre\\ for Astronomy
and Astrophysics, Pune, India}
\author{Sunandan Gangopadhyay}
\email{sunandan@bose.res.in, sunandan.gangopadhyay@gmail.com}
\affiliation{Department of Physics and Astrophysics, 
West Bengal State University,\\
Barasat, North 24 Paraganas,\\ West Bengal, India\\ }
\affiliation{Visiting Associate in S.N.Bose National Centre \\for
Basic Sciences, Kolkata, India}
\date{\today}


\begin{abstract}
{\noindent We consider the quantum dynamics of a test particle 
in noncommutative space under the 
influence of linearized gravitational 
waves in the long wave-length and low-velocity limit. 
A prescription for quantizing the classical 
Hamiltonian for the interaction of 
gravitational wave with matter in noncommutative space 
is proposed. The Hamiltonian (and hence
the system) is then exactly solved by using standard algebraic methods. 
The solutions show prominent signatures of the noncommutative nature
of space. Computation of the 
expectation value of the particle's position 
reveals the inherent quantum nature of spacetime noncommutativity.}
\end{abstract}


\pacs{11.10.Nx, 03.65.Ta, 11.10.Ef, 04.30.Nk, 42.50.Dv}
\maketitle

\noindent Gravitational waves (GWs) are tiny 
disturbances in spacetime. Their effect on 
matter is expected to be measured by the 
relative optical phase shift between the 
light paths in two perpendicular km-length 
arm cavities caused by the tiny displacement 
of two test mass mirrors (hung at the end of each cavity) 
induced by passing GWs. The typical amplitude of 
GWs $h \sim \frac{\delta L}{L}$ emitted by binary 
systems in the VIRGO cluster of galaxies, 
at a distance $\sim 20 {\rm Mpc}$, at $100 {\rm Hz}$ 
which the ground based Laser-interferometric 
GW detecters like LIGO \cite{abramovici}, VIRGO \cite{caron}, 
GEO \cite{luck} TAMA \cite{ando} etc. 
are designed to probe is $\sim 10^{-21}$. 
Since the cavity arm length $L \sim 1 {\rm km}$, 
$\delta L$ will be $\sim 10^{-18}{\rm m}$. 
It is, therefore, at the quantum mechanical 
level that experimental evidence for the GWs is 
likely to appear \cite{Caves}. 

\noindent Interestingly, in recent developments 
of noncommutative (NC) quantum mechanics and 
NC quantum field theory, where the coordinates 
$x^{\mu}$ satisfy the NC algebra
\begin{equation}
\left[x^{\mu}, x^{\nu}\right] = i \theta^{\mu \nu}
\label{ncgometry}
\end{equation}
the upperbounds on various NC parameters appearing 
in the literature \cite{carol, bert0, RB, ani, stern, mpr, cst} 
are quite close to this length scale. A wide range of 
theories have been constructed in a NC framework 
including various gauge theories \cite{sz}, gravity 
\cite{grav} and even encompassing certain possible 
phenomenological consequences \cite{jabbari1, rs5, rs7, rs9, rs10}. 
The upperbound on the value of the coordinate commutator 
$\theta^{ij}$ found in \cite{carol} is $\lesssim \left(10 {\rm TeV}\right)^{-2}$ which corresponds to $4 \times 10^{-40} {\rm m}^{2}$ for $\hbar$$=$$c$$=$$1$. Whereas such upperbounds on time-space NC parameter $\theta^{0i}$ is $\lesssim 9.51\times 10^{-18} {\rm m}^{2}$. However, recent studies in NC quantum mechanics revealed that the NC parameter associated with different particles are not same \cite{pmho} and this bound could be as high as $\theta \lesssim \left(4 {\rm GeV}\right)^{-2} - \left(30 {\rm MeV}\right)^{-2}$ \cite{stern}. These upperbounds correspond to the length scale $\sim 10^{-20} {\rm m} - 10^{-17} {\rm m}$.

\noindent With the prospect of the direct detection of GWs of such tiny amplitude as $\sim 10^{-18} $ 
in the near future, a sharp possibility of detecting the NC structure of spacetime would be in the GW detection experiments. Therefore, analyzing the interplay of classical GWs with a test particle in a {\it NC quantum mechanical} framework becomes important in its own right as it may predict some detectable signature of noncommutativity. 


\noindent In light of these observations, we would like to construct the quantum mechanics of a test particle in NC spacetime, interacting with a linearised GW in the long wave-length and low velocity limit. 
Since it has been demonstrated in various formulations of NC general relativity \cite{grav}, \cite{banerjee11} 
that any NC correction in the gravity sector 
is second order in the NC parameter, therefore, in a first order theory in NC space, the linearised GW remains unaltered by NC effects and any NC correction appearing in the system will be through the particle sector only. 
This is true not only with the canonical (i.e. constant) form of 
noncommutativity but also for the Lie-algebraic NC spacetime \cite{banerjee11}.
We shall incorporate the NC effect in the problem by writing the NC Hamiltonian and then reexpressing it in terms of the commutative coordinates and their momenta by the well known Bopp-shift transformations \cite{stern, cst}. Then we shall quantize the system following $\cite{speli}$. As a first step towards the formulation of a NC quantum mechanics of GW interacting with a test particle, we shall concentrate on the simplest of the GWs, namely the linearly polarized ones in this analysis.  

\noindent In a linearised theory of gravity, once we choose the transverse-traceless (TT) gauge, 
all the gauge redundancies of the theory get removed and the GW is characterised by the only non-zero components \cite{Magg}
$h_{11} = -h_{22},$ and $ h_{12} = h_{21}$, called the $+$ and $\times$ polarisation respectively.
The only non-trivial components of the curvature tensor in TT-gauge are\footnote{As usual, latin indices run from $1-3$.}  
\begin{equation}
{R^j}_{0,k0} = - \frac{\partial \Gamma^j_{0k}}{\partial t} = -\frac{1}{2}\frac{\partial^2h_{jk}}{\partial t^2}\>
\label{e4}
\end{equation}
and the geodesic deviation equation in the proper detector frame becomes \cite{Magg} 
\begin{equation}
m\frac{d^2 {x}^{j}}{dt^2} = - m{R^j}_{0,k0} {x}^{k}  \>
\label{e5}
\end{equation}
which governs the response of a scalar spin-zero test particle to the passage of a GW. Here ${x}^{j}$ is the proper distance of the test particle from the origin
and $m$ is its mass. For convenience, we have assumed that GW detectors can be reasonably isolated so that any external forces other than the GW interaction are negligible, i.e. the particles are considered to be free otherwise. Also, the GW is treated as an external classical field.
Eq.(\ref{e5}) can be used to describe the evolution of proper distance in TT-gauge frame as long as the spacial velocities involved are non-relativistic. Also, $|{x}^{j}|$ has to be much smaller than the typical scale over which the gravitational field changes substantially, i.e. the reduced wavelength $\frac{\lambda}{2\pi}$ of GW. This situation is referred to as the \textit{small-velocity and long wavelength limit}. Thus, with Eq.(\ref{e5}) we can analyze the interaction of GW with a detector \footnote{Note that this condition is satisfied by resonant bar detectors and earth bound interferometers but not by the proposed space-borne interferometers such as LISA \cite{lisa} or by the Doppler tracking of spacecraft.} which has a characteristic linear size $L \ll \frac{\lambda}{2\pi}$. 

\noindent The Lagrangian for the system, whose time evolution is described by Eq.~$(\ref{e5})$, can be written, upto a total derivative term\cite{speli} as 
\begin{equation}
{\cal L} = \frac{1}{2} m\dot {x}^2 - m{\Gamma^j}_{0k}\dot {x}_{j} {x}^{k} \>.
\label{e8}
\end{equation}
The canonical momentum corresponding to ${x}_{j}$ is ${p}_{j} = m\dot {x}_{j} - m \Gamma^j_{0k} {x}^{k}$ and the Hamitonian becomes
\begin{equation}
{H} = \frac{1}{2m}\left({p}_{j} + m \Gamma^j_{0k} {x}^{k}\right)^2 \>.
\label{e9}
\end{equation}
If we take GWs propagating along the $z$-axis, then due to the transverse natureof GWs, ${\Gamma^j}_{0k}$ has non-zero components only in the ${x}-{y}$ plane and the particle undergoes free motion along the $z$-direction. Hence we can essentially focus on the planar motion of the particle.

\noindent To ``quantize" this system in the NC plane, we now replace $x^{j}$ and $p_{j}$ in the above Hamiltonian by operators ${\hat x}^{j}$ and ${\hat p}_{j}$ satisfying the NC Heisenberg algebra\footnote{We would like to mention that
it is possible to shift the noncommutativity from the coordinates to the
momenta leading to a dual description as shown in the literature 
\cite{banerjeempla}.} 
\begin{eqnarray}
\left[{\hat x}_{i}, {\hat p}_{j}\right] = i\hbar \delta_{ij} \>, \quad 
\left[{\hat x}_{i}, {\hat x}_{j}\right] = i \theta \epsilon_{ij} \>,\quad 
\left[{\hat p}_{i}, {\hat p}_{j}\right] = 0\>.
\label{e9a}
\end{eqnarray}
It is well known that this can be mapped to the standard 
$\left( \theta = 0 \right)$ Heisenberg algebra spanned by $X_{i}$ and $P_{j}$ using \cite{cst, stern}
\begin{eqnarray}
{\hat x}_{i} = X_{i} - \frac{1}{2 \hbar} \theta \epsilon_{ij} P_{j}\>, \quad {\hat p}_{i} = P_{i} \>.
\label{e9b}
\end{eqnarray}
Using the traceless property of the GW and rewriting the NC version of Eq.$(\ref{e9})$ in terms of the operators $X_{i}$ and $P_{j}$, we obtain
\begin{equation}
{\hat H} = \frac{ P_{j}{}^{2}}{2m} + \Gamma^j_{0k} X_{j} P_{k} - \frac{\theta }{2 \hbar} \epsilon_{jm} P_{m} P_{k}  \Gamma^j_{0k} \>. 
\label{e12}
\end{equation}
Observe that a direct coupling between the GW and spatial noncommutativity appears here. Since we are dealing with linearized gravity, 
a term quadratic in $\Gamma$ has also been neglected in Eq.$(\ref{e12})$. 

\noindent It is to be noted that the above 
Hamiltonian can be obtained from the following
Lagrangian (which can be computed from the Hamiltonian (\ref{e12}) 
by an inverse Legendre transformation)
\begin{equation}
\hat{\mathcal{L}} = \frac{m}{2}\left[(\dot{X}_{i}^{2}-2{\Gamma^{j}}_{0i}
\dot{X}_{i}X_{j})+\frac{\theta m}{\hbar}\epsilon_{jk}{\Gamma^{j}}_{0i}
\dot{X}_{i}\dot{X}_{k}\right]~.
\label{lag12}
\end{equation}
This is reassuring since we do have both the Lagrangian and the Hamiltonian
of the problem in hand as in the commutative ($\theta =0$) case. 
Further, the above form of the
Lagrangian can be obtained from the noncommutative version of the
Lagrangian (\ref{e8}) by making the following Bopp shift 
\begin{eqnarray}
{\hat x}_{i} = X_{i} - \frac{1}{2 \hbar} \theta \epsilon_{ij} P_{j}\>, 
\quad \hat{\dot{x}}_{i} = \dot{X}_{i} 
\label{bopp_shift_lag}
\end{eqnarray}
and then putting $P_{i}=m\dot{X}_{i}$ since both the quadratic terms
in $\theta$ and $\Gamma$ can be neglected. This can be regarded 
as the Lagrangian version of the Bopp shift (\ref{e9b}) applied at the
Hamiltonian level .

\noindent Defining raising and lowering operators
\begin{eqnarray}
X_j &=& \left({\hbar\over 2m\varpi}\right)^{1/2}\left(a_j+a_j^\dagger\right)\>\label{e15a} \\
P_j &=& -i\left({\hbar m\varpi\over 2}\right)^{1/2} \left(a_j-a_j^\dagger\right)\>
\label{e15}
\end{eqnarray}
where $\varpi$ is determined from the initial uncertainty in either the position or the momentum of the particle, we write the Hamiltonian (\ref{e12}) as 
\begin{eqnarray}
{\hat H} &=& \frac{\hbar\varpi}{4}\left(2 a_j^\dagger a_j + 1 - a_j^2 - {a_j^\dagger}^2\right) - \frac{i\hbar}{4} \dot h_{jk} \left(a_j a_k - a_j^\dagger a_k^\dagger\right) \nonumber \\
&& + \frac{m \varpi \theta}{8} \epsilon_{jm} {\dot h}_{jk}  \left(a_{m}a_{k}  - a_{m}a_{k}^\dagger + C.C \right)\>
\label{e16}
\end{eqnarray}
where C.C means complex conjugate. Working in the Heisenberg representation, the time evolution of $a_{j}(t)$ is given by 
\begin{eqnarray}
\frac{da_j}{dt} &=& -i\frac{\varpi}{2}(a_j-a^\dagger_j) + \frac{1}{2}\dot
		   h_{jk}a^\dagger_k \nonumber \\
&& + \frac{i m \varpi \theta}{8 \hbar} \left(\epsilon_{lj} {\dot h}_{lk} + \epsilon_{lk} {\dot h}_{lj}\right)\left(a_{k} - a^\dagger_{k}\right)\>
\label{e17}
\end{eqnarray}
and that of $a_{j}^{\dagger}(t)$ is the C.C of Eq.$(\ref{e17})$. Next, noting that the raising and lowering operators must satisfy the commutation relations
\begin{equation}
[a_j(t),a_k(t)] = 0\>,\qquad [a_j(t),a^\dagger_k(t)] = \delta_{jk}\>
\label{e18}
\end{equation}
we write them in terms of $a_j(0)$ and $a_j^{\dagger}(0)$, the free operators at $t=0$, by the time-dependent Bogoluibov transformations
\begin{eqnarray}
a_j(t) &=& u_{jk} a_k(0) + v_{jk}a^\dagger_k(0)\>
\nonumber \\
a_j^\dagger(t) &=& a_k^\dagger(0)\bar u_{kj}  + a_k(0)\bar v_{kj}\>
\label{e19}
\end{eqnarray}
where the bar denotes the C.C and $u_{jk}$ and $v_{jk}$ are the generalized Bogoluibov coefficients. They are
$2\times 2$ complex matrices which, due to eq.~$(\ref{e18})$, must satisfy
$uv^{T}=u^{T}v\>,\> u u^\dagger - v v^\dagger = I,$
written in matrix form where $T$ denotes transpose, $\dagger$ denotes conjugate transpose and $I$ is the identity
matrix. Since $a_j(t = 0) = a_j(0)$, $u_{jk}$ and $v_{jk}$ have the boundary conditions $u_{jk}(0)= I $ and
$v_{jk}(0) = 0$. Then, from eq.$(\ref{e17})$ and its C.C we get the equations of motions in terms of $\xi = u + v^\dagger$ and $\zeta = u - v^\dagger$:
\begin{eqnarray}
\frac{d \xi_{jk}}{dt} &=& -i\varpi \zeta_{jk} + \frac{{\dot h}_{jl}}{2}\xi_{lk} +\Theta_{jl} \zeta_{lk}\> 
\label{e21a} \\
\frac{d \zeta_{jk}}{dt} &=& -\frac{1}{2}{\dot h}_{jl}\zeta_{lk}\> 
\label{e21b}
\end{eqnarray}
where $\Theta_{jl}$ is the new term reflecting the interplay of noncommutativity with GW
\begin{eqnarray}\Theta_{jl} = \frac{i m \varpi \theta}{4 \hbar}\left({\dot h}_{jm} \epsilon_{ml} - \epsilon_{jm} {\dot h}_{ml}\right)\>.
\label{e21ab}
\end{eqnarray}
Eq(s).~$(\ref{e21a}, \ref{e21b})$ are difficult to solve analytically for general $h_{jk}$. However, our gole, in the present letter, is to investigate {\it to what extent spatial noncommutativity affects the interaction of GWs with spin-zero test particle} in the simplest of settings. Therefore we shall solve Eq(s).~$(\ref{e21a}, \ref{e21b})$ for the special case of linearly polarized GWs. 

\noindent In the two-dimensional plane, the GW, which is a $2\times 2$ matrix $h_{jk}$, is most conveniently written in terms of the Pauli spin matrices as 
\begin{equation}
h_{jk} \left(t\right) = 2f(t) \left(\varepsilon_{\times}\sigma^1_{jk} + \varepsilon_{+}\sigma^3_{jk}\right) = 2f(t)\varepsilon_A\sigma^A_{jk}\>.
\label{e13}
\end{equation}
Note that the index $A$ runs from $1-3$, however, no contribution from $\sigma^2$ is included. $2f(t)$ is the amplitude of the GW whereas $\varepsilon_{\times} \left(t \right)$ and $\varepsilon_{+} \left( t \right)$  represent the two possible polarization states of the GW and satisfy the condition
$\varepsilon_{\times}^2+\varepsilon_{+}^2 = 1$
for all $t$. In case of linearly polarized GWs however, the polarization states $\varepsilon_{A}$ are independent of time and $f(t)$ is arbitrary. To set a suitable boundary condition we shall assume that the GW hits the particle at $t=0$ so that \begin{equation}
f(t)=0 \>, \quad {\rm for} \ t \le 0.
\label{bc}
\end{equation}

\noindent We now move on to solve Eq.$(\ref{e21b})$ by noting that any $2\times 2$ complex matrix can be written as a linear combination of the Pauli spin matrices and identity matrix. Hence we make the ansatz : 
\begin{equation}
\zeta_{jk}\left(t \right) = A \left(t \right) I_{jk} + B\left(t \right) \varepsilon_{A} \sigma^{A}_{jk}\>. 
\label{form1}
\end{equation} 
Substituting for $h_{jk}$ and $\zeta_{jk}$ from Eq.$(\ref{e13})$ and Eq.$(\ref{form1})$ respectively in Eq.$(\ref{e21b})$ and comparing the coefficients of $I$ and $\varepsilon_{A} \sigma^{A}$, we get first order differential equations for $A$ and $B$ which can be readily integrated to obtain 
\begin{equation}
\zeta_{jk} = \cosh \left[f\left(t\right)\right] I_{jk} -\sinh \left[f\left(t\right)\right]\varepsilon_{A} \sigma^{A}_{jk}\>
\label{zeta}
\end{equation}
once we have used the boundary conditions Eq.$(\ref{bc})$.

\noindent Now that we have the solution for $\zeta$, we can take a closer look at the NC correction term in Eq.$(\ref{e21ab})$, before proceeding to solve Eq.$(\ref{e21a})$. Computation shows that the coupling between noncommutativity and GW not only deforms the effects of the two polarization states $\varepsilon_{\times}\sigma^1_{jk}$ and $\varepsilon_{+}\sigma^3_{jk}$ present in the GW but also generates a new term proportional to the Pauli spin matrix $\sigma^{2}_{jk}$ :
\begin{eqnarray}
\Theta_{jl} \zeta_{lk} = \frac{m \varpi \theta}{\hbar}{\dot f}\left( t \right)\left\{ i A \left(\varepsilon_{+}\sigma^{1} - \varepsilon_{\times}\sigma^{3}\right)_{jk} + B \sigma^{2}_{jk} \right\}\>.
\label{cor}
\end{eqnarray}  
Hence, as a trial solution we take a more general form for $\xi_{jk}$ that accomodates all the Pauli matrices along with the identity matrix
\begin{eqnarray}
\xi_{jk} \left(t \right)=  C I_{jk} + D_{1} \sigma^{1}_{jk} + D_{2} \sigma^{2}_{jk} + D_{3} \sigma^{3}_{jk}\>.
\label{form2}
\end{eqnarray} 
Substituting for $h_{jk}$, $\zeta_{jk}$ and $\xi_{jk}$ from Eq(s).$(\ref{e13})$, $(\ref{form1})$ and $(\ref{form2})$ respectively in Eq.$(\ref{e21a})$ and comparing the coefficients of $I$ and $\sigma^{A}$, we get the following set of first order differential equations 
\begin{eqnarray}
{\dot C} &=& -i \varpi A + {\dot f}\left(D_{1} \varepsilon_{\times} + D_{3} \varepsilon_{+}\right)\>\nonumber \\
{\dot D}_{1} &=& -i B \varpi \varepsilon_{\times} + {\dot f}\left( C \varepsilon_{\times} - i D_{2} \varepsilon_{+} + \frac{i m \varpi \theta}{\hbar} A \varepsilon_{+}\right)\> 
\nonumber \\
{\dot D_{2}} &=& i{\dot f}\left( D_{1} \varepsilon_{+} - D_{3} \varepsilon_{\times}\right) + \frac{m \varpi \theta}{\hbar}  {\dot f} B\> \nonumber \\
{\dot D}_{3} &=& -i B \varpi \varepsilon_{+} + {\dot f}\left( C \varepsilon_{+} + i D_{2} \varepsilon_{\times} - \frac{i m \varpi \theta}{\hbar} A \varepsilon_{\times}\right)\>.
\label{set1}
\end{eqnarray}
With the recombination 
\begin{eqnarray}
G = D_{1} \varepsilon_{\times} + D_{3} \varepsilon_{+} \quad, \quad Q = D_{1} \varepsilon_{+} - D_{3} \varepsilon_{\times}\>
\label{combo}
\end{eqnarray}
we recast the set of Eq(s).$(\ref{set1})$ in the following form 
\begin{eqnarray}
{\dot C} &=& -i \varpi A + {\dot f} G \quad , \quad
{\dot G} = -i \varpi B + {\dot f} C \label{set2a} \\
{\dot D_{2}} &=& i{\dot f} Q + \frac{m \varpi \theta}{\hbar}{\dot f} B\>, \
{\dot Q} =  - {\dot f} D_{2} + \frac{m \varpi \theta}{\hbar} i{\dot f} A
\label{set2b}
\end{eqnarray}
Integrating Eq(s).$(\ref{set2a})$ and $(\ref{set2b})$ and imposing the boundary condition Eq.$(\ref{bc})$ we get 
\begin{eqnarray*}
C &=&  \cosh f(t) - \frac{i \varpi}{2} F_{+} \left( t \right)  \>; \quad G = -\frac{i \varpi}{2} F_{-}\left( t \right)
; \\
D_{2} &=& -\frac{m \varpi \theta}{\hbar} f\left( t\right) \sinh\left[f\left( t\right)\right] \>
; \\
Q &=& -\frac{i m \varpi \theta}{\hbar} f\left( t\right) \cosh\left[f\left( t\right)\right]
\>;\\ 
F_{\pm}\left( t \right) &=& \left[e^{f\left( t\right)} \int^{t}_{0} dt^{\prime} e^{-2f\left( t^{\prime}\right)} \pm e^{-f\left( t\right)} \int^{t}_{0} dt^{\prime} e^{2f\left( t^{\prime}\right)}\right]
\>. 
\end{eqnarray*}
Inverting the set of equations in $(\ref{combo})$, we can express $D_{1}$ and $D_{3}$ in terms of $G$ and $Q$ 
. Combining the expressions for $C$, $D_{1}$, $D_{2}$ and $D_{3}$ we can write the solution for $\xi$ using Eq.(s) $(\ref{form2})$ which along with $\zeta$ in Eq. $(\ref{zeta})$, give $u_{l}$ and $v_{l}$, where $l$ stands for linear polarization : 
\begin{widetext}
\begin{eqnarray}
u_{l}\left(t\right) &=& I \left[\cosh f - i\frac{\varpi}{4}F_{+}\right] - \frac{i\varpi}{4} \left( \epsilon_{\times}\sigma^{1} + \epsilon_{+} \sigma^{3} \right)F_{-} + \left(\frac{ m \varpi \theta}{2 \hbar}\right) f\left\{i\left( \epsilon_{+} \sigma^{1} - \epsilon_{\times} \sigma^{3} \right)\cosh f + \sigma^{2} \sinh f \right\}\>, \label{u}\\ 
v_{l}\left(t\right) &=& I \frac{i\varpi}{4} F_{+} + \left( \epsilon_{\times}\sigma^{1} + \epsilon_{+} \sigma^{3} \right) \left[\sinh f + i\frac{\varpi}{4} F_{-}\right] - \left(\frac{ m \varpi \theta}{2 \hbar}\right) f\left\{i\left(\epsilon_{+} \sigma^{1} - \epsilon_{\times} \sigma^{3} \right)\cosh f + \sigma^{2} \sinh f \right\} \>.
\label{v}
\end{eqnarray}
\end{widetext}
\noindent The system has now been essentially solved for the linearly polarized GWs once one specifies the initial expectation values of the particle position and momentum and also the initial uncertainty in either the position or the momentum of the particle to determine $\varpi$. For example, one can assume the initial wave-function \footnote{$x$ is the eigen value of the position operator $X$ and $\vec{p}$ is the eigenvalue of the momentum operator $\vec{P}$ and $l_{0} = \sqrt{\frac{\hbar}{m \varpi}}$ sets the length scale.} of the particle to be a Gaussian wave-packet 
\begin{eqnarray}
\psi\left(x, y \right) = \frac{1}{\sqrt{\pi}l_{0}} e^{ -\frac{1}{2 l_{0}^{2}}\left[\left(x-x_{0}\right)^{2} + \left(y-y_{0}\right)^{2}\right]} e^{\frac{i \vec p .\vec r}{\hbar}}
\label{psi}
\end{eqnarray}
located at $\left(x_{0}, y_{0}\right)$ and moving with momentum $\vec{p}_{0} = \left(p_{x_{0}}, p_{y_{0}}\right)$ when the GW just hits the system at time $t=0$. This will give uncertainties in momentum and position as $\sqrt{\frac{\hbar m \varpi}{2}}$ and $\sqrt{\frac{\hbar }{2m \varpi}}$ respectively. Either can be used to compute $\varpi$. From the initial position and momentum expectation values, 
i.e. $\langle\vec{r}_{0} \rangle= \left(x_{0}, y_{0}\right)$ and $\langle\vec{p}_{0}\rangle = \left(p_{x_{0}}, p_{y_{0}}\right)$, we get the raising and lowering operator at time $t = 0$. 
We then use  Eq(s) $(\ref{e19})$, $(\ref{u})$ and $(\ref{v})$ 
to find $a_{j}\left( t \right)$ and $a_{j}^{\dagger}\left( t \right)$ 
at a general time. This in turn gives us the expectation 
value of the position of the particle at any arbitrary time $t$
\begin{eqnarray}
\langle x\left(t\right)\rangle &=& \cosh f(t)x_{0} + 
\left(\epsilon_{+}x_{0} + \epsilon_{\times}y_{0} \right)
\sinh f(t) + \left(F_{+} \right. \nonumber\\
&+& \left. \epsilon_{+} F_{-}\right) \frac{p_{x_0}}{2 m} + 
\epsilon_{\times}F_{-}\frac{p_{y_0}}{2m} + 
\frac{\theta p_{x_0}}{\hbar} \epsilon_{\times}f(t)\cosh f(t)\nonumber\\
&& +  \frac{\theta p_{y_0}}{\hbar}\left[ f(t)\sinh f(t) - 
\epsilon_{+} f(t) \cosh f(t) \right] 
\label{x}\\
\langle y \left(t\right)\rangle &=& \cosh f(t)y_{0} + 
\left(\epsilon_{\times}x_{0} - 
\epsilon_{+}y_{0} \right) \sinh f(t) + \left(F_{+} 
\right. \nonumber\\
&-& \left. \epsilon_{+} F_{-}\right)\frac{p_{y_0}}{2 m} + 
\epsilon_{\times}F_{-}\frac{p_{x_0}}{2m} - 
\frac{\theta p_{y_0}}{\hbar} \epsilon_{\times} f(t)\cosh f(t) \nonumber\\
&& - \frac{\theta p_{x_0}}{\hbar}\left[f(t)\sinh f(t) + \epsilon_{+} f(t) 
\cosh f(t) \right] 
\label{y}
\end{eqnarray}
Interestingly, we find that the NC nature of spacetime 
affects only those particles which are in motion at time $t=0$. 
Further, the presence of $\frac{1}{\hbar}$ factor in the 
NC correction even after the computation of the expectation value 
indicates that the NC effect is 
inherently quantum mechanical in nature. 
To make the connection with the well known classical result, 
we first set $f\left(t\right) = f_{0}$ ($f_{0}$ being 
a small constant amplitude of GW) and $y_{0} = p_{y_{0}} = 0$. From Eq(s) 
(\ref{x}) and (\ref{y}), we can immidiately see 
\begin{eqnarray}
\langle x\left(t\right)\rangle &=& \left(x_{0} + \frac{p_{x0}t}{m}\right) + \epsilon_{+}f_{0} \left(x_{0} -  \frac{p_{x0}t}{m}\right) + \frac{\theta p_{x0}}{\hbar} \epsilon_{\times}f_{0} \nonumber\\
\label{x1}\\
\langle y\left(t\right)\rangle &=&  \epsilon_{\times}f_{0} \left(x_{0} -  \frac{p_{x0}t}{m}\right) - \frac{\theta p_{x0}}{\hbar} \left(\epsilon_{+} + f_{0}\right)f_{0}~. 
\label{y1}
\end{eqnarray}
Now putting $\theta  = 0$ gives us the well 
known classical result \cite{carol} in the low-velocity, 
long-wavelength limit. 
From the above expressions for the expectation values, we find that
the NC effect for a particle interacting 
with GW appears in the form of a product
of $\theta$ and $f_{0}$ and hence it
would be difficult to detect this effect as it is very small 
compared to the term showing the effect of the passing GW. 
This is due to the fact that noncommutativity has been introduced 
by Bopp-shifting the coordinates which couple with the $\Gamma$-term in the 
Hamiltonian and hence the NC correction in the 
Hamiltonian (\ref{e12}) contains a $\Gamma$-term. 
However, the present exercise gives 
us a substantial idea of what to expect if the same computational scheme 
is applied for a harmonic oscillator system. 
The NC effect is expected to be more prominent in this case
since along with the coupling 
term of GW with noncommutativity, there will also be 
NC corrections coming from the harmonic oscillator potential 
which may give rise to a NC effect of comparable 
magnitude with GW interaction. Interaction of GW with a 
Hydrogen atom in the NC setting will be more involved \cite{stern} 
and is yet another interesting case to study where a 
detectable NC effect should occur. 
Work in this direction is in progress and will be taken up 
in the subsequent papers. 

\section*{Acknowledgment}
\noindent The authors would like to acknowledge 
the hospitality of IIT Kanpur where a considerable part of this work was 
completed. The authors would also like to thank the referee for
useful comments. 


\end{document}